\documentclass[%
 reprint,
preprintnumbers,
nofootinbib,
 amsmath,amssymb,
 aps,
]{revtex4-2}

\usepackage{graphicx}
\usepackage{dcolumn}
\usepackage{bm}
\usepackage{xcolor}
\usepackage{hyperref,aas_macros}

\newcommand{\vk}{\mathbf{k}}
\newcommand{\done}{\hat{\delta}_g^{(1)}}
\newcommand{\dtwo}{\hat{\delta}_g^{(2)}}

\begin{document}

\preprint{MIT-CTP/6059}

\title{Sample Variance Cancellation for Future Spectroscopic Surveys}

\author{James M. Sullivan}
 \email{jms3@mit.edu}
 \thanks{Brinson Prize Fellow}
 \affiliation{
 Center for Theoretical Physics - a Leinweber Institute, Massachusetts Institute of Technology, Cambridge, MA 02139, USA}
\author{Martin White}%
\email{mwhite@berkeley.edu}
\affiliation{Department of Physics, University of California, Berkeley, CA 94720, USA}
\affiliation{Berkeley Center for Cosmological Physics, UC Berkeley, CA 94720, USA}
\affiliation{Lawrence Berkeley National Laboratory, One Cyclotron Road, Berkeley, CA 94720, USA
}

\date{\today}

\begin{abstract}
High-redshift spectroscopic galaxy surveys will be the scientific engines of the next generation of large-scale structure cosmology.
The clustering signal of high-redshift, star-forming Lyman-$\alpha$ emitters (LAEs) will be of key importance for obtaining high-redshift constraints on the growth of structure and redshift-space distortions.
The complex radiative transfer (RT) of Lyman-$\alpha$ photons alters the symmetry group respected by the overdensity field constructed from these galaxies, and so the observed large-scale clustering of LAEs may have an angular dependence that differs significantly from that of linear theory, possibly biasing inference of cosmological parameters.
While such an effect has been seen in simulations, its amplitude in nature and its detailed form remains unclear.
In the restricted context of a linear, Gaussian model, we outline a procedure for pinning down the type and amplitude of such changes in angular dependence on large scales due to unknown RT or a more general unmodeled angular effect
in the
hypothetical scenario in which an observer is presented with LAE data containing such an effect.
We show that if a second tracer without the modified angular dependence is available for cross correlation with the LAEs at the same redshifts (e.g., Lyman-break galaxies), then, with a high-redshift survey of modest size, it is possible to rapidly identify: 1) the presence of a nontrivial angular functional form of radiative transfer (by a conditional field-level realization), 2) the functional form itself (with an optimal filter that we derive), and 3) the value of its amplitude with an uncertainty (via an adaptation of the standard quadratic estimator).
Such sample-variance-cancellation strategies therefore provide a statistical solution to unknown astrophysical or systematic angular clustering dependence, including from RT.
\end{abstract}

\maketitle


\section{Introduction \label{sec:intro}}

The study of large-scale structure with large galaxy surveys has entered a new era of precision with the ongoing DESI \cite{DESI16-I,DESI16-II}, Euclid \cite{Euclid11,Euclid18} and PFS \cite{PFS} surveys dramatically increasing the volume and number of galaxies with spectroscopic redshifts.
The next generation of surveys aims to further the precision of galaxy clustering by pushing the high-redshift frontier, and these surveys are currently in the planning stages \cite{Bacon24,Beseuner25}, targeting primarily star-forming galaxies.
With increasing statistical precision comes the need to ensure that systematic errors remain controlled.
Precision galaxy clustering measurements are largely unexplored in the high-redshift regime targeted by future spectroscopic galaxy surveys.
It is therefore natural to expect the presence of systematic effects in these measurements that have not been previously considered for these new tracers.

Galaxies that are selected by their line emission may result in a galaxy density field that depends explicitly on local velocity and density gradients projected along the line of sight.
For most spectral lines this effect is irrelevant, but due to the ubiquity of hydrogen in the interstellar medium (ISM) and intergalactic medium (IGM) as well as the complicated radiative transfer of the Lyman-$\alpha$ (Ly$\alpha$) line, galaxy samples selected by this line may exhibit anisotropic clustering beyond the usual Kaiser effect \cite{Kaiser1987}.
As their name implies, Lyman-$\alpha$ emitting galaxies (LAEs) will be selected in just this way, and therefore their clustering may exhibit a dependence on these extra terms.
These galaxies are of special interest for future spectroscopic surveys, as they have a (relatively) low bias and high number density, putting them in the preferred regime to probe the majority of leading primordial physical effects that are largely informed by linear modes. 
At the linear level, RT effects would show up in the bias expansion for the galaxy density field as
\begin{equation}
    \label{eqn:delta_g_linear}
    \delta_g = b\delta_m + b_\theta \theta
\end{equation}
with $\delta_m$ the matter overdensity and $\theta$ its velocity divergence.
In the absence of RT, $b_\theta\equiv f \approx 1$.
Failure to account for this effect on the observed clustering of such galaxies may lead to biased cosmological inference.

Hydrodynamical simulations with astrophysical models of Ly$\alpha$ radiative transfer have attempted to quantify the size of this effect, beginning with Ref.~\cite{Zheng11}, who illustrated its impact for the first time.
While Ref.~\cite{Zheng11} found the impact of hydrogen ISM/IGM geometry on clustering to be large, subsequent investigation by several groups found a smaller impact on clustering and a significant numerical resolution dependence \cite{Behrens18,Behrens:2018_rt2,Gurung19,Momose21,Khoraminezhad:2025_lae_rt}.
However, with the same model as Ref.~\cite{Gurung19}, Ref.~\cite{Gurung-Lopez:2020_rt1_2} found that, especially at higher redshift, the details of the ISM and IGM modeling have a significant impact on the correlation of observed LAEs with LOS velocity gradients, though the effect is numerically small for the main redshifts of interest for upcoming LAE samples.

Since no hydrodynamical simulation can accommodate the enormous range of scales required to simultaneously resolve the ISM/CGM and cosmologically relevant scales, a fully self-consistent model of the impact of the ISM/CGM/IGM on LAE selection will almost certainly remain out of reach in the foreseeable future.
In addition, the lack of consensus in the literature due to strong astrophysical model dependence coupled with the limited constraints from observed data on Ly$\alpha$ radiative transfer strongly motivates an observational treatment of this issue.

Thankfully, the multi-tracer formalism \cite[e.g.,][]{Seljak_mt_png,McDonaldSeljak:2009,Abramo2016:mt_p_qe,Rubiola:2025_wf_mt_grdm,TRVA2022:multittracer,Rubira2025:mt} provides a data-driven way of determining the presence of clustering components due to Ly$\alpha$ radiative transfer.
Because this astrophysical effect is a property that can be written as a ratio of power spectra, it can be determined without sample variance, as is the case for the growth rate in redshift space distortions or, famously, the amplitude of local primordial non-Gaussianity \cite[e.g.,][]{McDonaldSeljak:2009,Seljak_mt_png,Sullivan2023:concentrate_mt,BarreiraKrause2023:fnl_multi,Fondi2024:MT}.
More pointedly, while a single power spectrum measurement in a $k$ bin can only be estimated with error proportional to the inverse of $\sqrt{N_\mathrm{mode}(k)}$, the comparison of two fields that sample the same underlying cosmological modes do not suffer from this limitation.

The goal of this paper is to offer a solution to a hypothetical LAE clustering observer who wants to know if their data has evidence of unmodeled RT\footnote{We will often use ``RT'' as a motivating example, but the techniques apply to any unmodeled clustering angular dependence.}.
We showcase several tools using multiple tracers that can be used to quantify and mitigate the impact of unknown RT effects on clustering measurements.
Each tool will be more or less appropriate depending on the quality of knowledge at hand -- though all rely on Gaussianity of the tracer fields, which we expect to be an excellent approximation for a wide range of scales at the high redshifts relevant for future spectroscopic surveys.
If one has little information on the form of the RT effect or strongly suspects that the model for the $\mu$ dependence is wrong, one can perform a simple model misspecification test using a second tracer with known $\mu-$dependence (e.g.\ with no RT). 
After identifying at least the presence of an unknown RT, one can then determine its functional form by applying ``multi-tracer weights'' to the field computed under a variety of models for the $\mu-$dependence of the RT effects\footnote{The weights will be sub-optimal for the wrong model, but will only increase uncertainty.}.
Finally, once a decently approximating RT clustering model has been found, a stronger assumption of a particular $\mu-$dependence for the RT effect can be made and the unknown amplitude can be constrained 
optimally via a quadratic estimator.

We provide an overview of the relevant quantities that arise when considering tracer cross-spectra in Section~\ref{sec:orientation} before illustrating how unknown $\mu-$dependent clustering affects forecasted knowledge of RT when the RT model is known, and the benefit of including multiple tracers.
We then present a sequence of multi-tracer RT-diagnostic tools when the model is unknown: conditional density RT model misspecification (Section~\ref{sec:cond_dens}), the optimal weights under a given RT model
(Section~\ref{sec:opt_filt}) and a quadratic estimator assuming a well-approximating fiducial RT model (Section~\ref{sec:qe}) before concluding in Section~\ref{sec:conc}.

\section{Orientation \label{sec:orientation}}

On large scales, where linear theory is valid, and holding fixed the cosmology, a measurement of the power spectrum monopole, $P_0$, is sufficient to determine the large-scale bias, $b$.  Knowing $b$ then provides a prediction for the quadrupole, $P_2$.  If the prediction doesn't match our observations, either our assumed cosmology is wrong, linear theory doesn't hold or we have evidence for additional physics.  Unfortunately this method can be quite noisy, since on the scales where linear theory is expected to hold sample variance from finite survey volume is large.  Fortunately the presence of a second tracer, assumed to be immune to the effects under consideration, allows us to do much better than this first attempt.

To orient the reader let us consider a Gaussian matter density field, $\delta_m(\mathbf{k})$, that is being traced by galaxies as $\delta_g=b\delta_m+\epsilon$.  Here $b$ is a scale-independent, linear bias parameter and $\epsilon$ is a `noise' that is uncorrelated with $\delta_m$.  At first we will work in real space, i.e.\ neglecting redshift-space distortions.  Defining $P_\epsilon=\langle\epsilon^2\rangle$ and $P_g=b^2P_m=b^2\langle\delta_m^2\rangle$
\begin{equation}
    r_{gm}^2
    \equiv \frac{\langle\delta_g\delta_m\rangle^2}{\langle\delta_g^2\rangle\,\langle\delta_m^2\rangle}
    = \frac{P_g}{P_g+P_\epsilon}
\end{equation}
which we recognize as the well-known relation between the Wiener filter and linear regression \footnote{We neglect all Dirac $\delta^{(D)}$ factors on power spectra when taking expectations.}.  A common noise source in galaxy surveys is shot-noise for which the Poisson value is $P_\epsilon=\bar{n}^{-1}$.  In this case $r_{gm}^2=\bar{n}P_g/[1+\bar{n}P_g]$.  When $\bar{n}P_g\gg 1$ the galaxy and matter fields are well correlated.  Other common sources of noise are non-linear bias (e.g.\ if $\delta_g\supset b_2[\delta_m^2-\langle\delta_m^2\rangle]$ then this is uncorrelated with $\delta_m$ and contributes `noise' $P_\epsilon\sim P\star P$) and non-linear evolution of the matter field.  At high redshift, shot-noise and non-linear bias dominate over non-linear evolution, while at low redshift the opposite occurs.  We shall be interested in the case where $\epsilon$ includes systematic errors.

If we knew $\delta_m$ then we could measure our `noise' by considering
\begin{equation}
    P_{\rm err} \equiv \mathrm{min}_\alpha \left\langle \left( \delta_g - \alpha \delta_m \right)^2 \right\rangle = (1-r_{gm}^2)P_g
\label{eqn:simple_Perr}
\end{equation}
If $r_{gm}$ is close to 1 this can be dramatically smaller than $P_g$, allowing us to probe even small deviations from our expectations.  This is often referred to as sample variance cancellation.

Of course, normally we don't have access to $\delta_m$.  However given a second galaxy field, which we believe to be free of the systematic or effect we are interested in testing, we can generalize Eq.~(\ref{eqn:simple_Perr}) to
\begin{equation}
    P_{\rm err} \equiv \mathrm{min}_\alpha \left\langle \left( \delta_g^{(1)} - \alpha \delta_g^{(2)} \right)^2 \right\rangle = (1-r_{12}^2)P^{(1)},
\end{equation}
for galaxy fields $\delta_g^{(1)},\delta_g^{(2)}$ with auto-power spectra $P^{(1)},P^{(2)}$, respectively.
Again when $r_{12}\approx 1$ this cancels a large part of the signal allowing us to see small levels of contamination.  Since typically the (fractional) error on the power spectrum scales as $V_{\rm surv}^{-1/2}$ in the sample variance limit, to obtain similar levels of sensitivity without sample variance cancellation would require a survey $(1-r_{12}^2)^{-2}$ larger \footnote{This is similar to the gain one obtains from control variates, e.g.\ ref.~\cite{Kokron22}, though since we work at the field level rather than the power spectrum the gain is even larger.}.

This additional information from the cross correlation of the two tracers can be seen as a larger effective volume of the survey.
Consider the signal over signal+noise ratio (or Wiener filter)
\begin{equation}
    \label{eqn:s_sn}
    R(\vk,z) = \langle s^\dagger C^{-1} s\rangle(\vk,z),
\end{equation}
where $s$ is a vector containing the desired signal and $C=V^{-1}\langle d^\dagger d\rangle$ with $d = s+\epsilon$ and noise power spectrum $\langle \epsilon^\dagger \epsilon\rangle$.
For a single tracer, we have $s = b\delta_m$ and
this expression (eqn.~\ref{eqn:s_sn}) reduces to 
\begin{equation}
\label{eqn:s_sn_1}
    R(\vk,z) \to V\left(\frac{\bar{n}P_g}{1+\bar{n}P_g}\right) .
\end{equation}
This expression is similar to the ``effective volume'', but for fields rather than power spectra.
For multiple tracers $\delta_g^{(1)},\delta_g^{(2)}$, we have signal $s = \{b^{(1)}\delta_m,b^{(2)}\delta_m\}$.
Assuming that the noise autocorrelation for each tracer is given by shot noise, and that we abuse notation to represent the cross-noise $\langle n^{(1)} n^{(2)}\rangle$ as $\bar{n}_{12}^{-1}$, eqn.~\ref{eqn:s_sn} becomes
\begin{align}
    \label{eqn:s_sn_2}
    R =
    \frac{V}{1-r_{12}^2}\Bigg[&\left(\frac{\bar{n}^{(1)}P^{(1)}_g}{1+\bar{n}^{(1)}P^{(1)}_g}\right) + \left(\frac{\bar{n}^{(2)}P^{(2)}_g}{1+\bar{n}^{(2)}P^{(2)}_g}\right) \nonumber \\
    &- 2 r_{12}^2\left(\frac{ \bar{n}_{12} P_g^{(12)}}{1 + \bar{n}_{12}P_{g}^{(12)} }\right)\Bigg]
\end{align}
where 
\begin{equation}
    \label{eqn:rcc_sn}
    r_{12}^{2} = \frac{\left(P_{g}^{(12)}+\bar{n}_{12}^{-1}\right)^2}{\left(P_{g}^{(1)}+\bar{n}_{1}^{-1}\right)\left(P_{g}^{(2)}+\bar{n}_{2}^{-1}\right)}.
\end{equation}
This expression for the cross-correlation coefficient reduces to the more common expression in the case of vanishing cross stochasticity (e.g.\ $\bar{n}_{12}^{-1}\to 0$).
While the expression in eqn.~\ref{eqn:s_sn_2} is more complicated than the single tracer case, we can already draw some intuitive conclusions from it in terms of volume rescaling.
We recover eqn.~\ref{eqn:s_sn_1} when the tracers are identical and obtain the naive sum of this expression for two tracers when the tracers are uncorrelated ($r_{12}\to 0$).
More interestingly, in the limit where the cross-noise is negligible\footnote{This turns out to be a good approximation for our use case, see discussion in Section~\ref{subsec:applications_Fisher_gmu2}.} and where the individual power spectra are detected with high signal-to-noise $\bar{n}_iP_i > 1$, we find the effective volume is boosted by a factor of $\left(r_{12}^2 R_1 + R_2/[1-r_{12}^2]\right)\to V\left(r_{12}^2  + 1/[1-r_{12}^2]\right)$, where $R_i$ correspond to eqn.~\ref{eqn:s_sn_1} for each tracer.  This can become very large as $r_{12}\to 1$.
As we will see later, the gain in sensitivity implied by this expression is often significant for upcoming spectroscopic surveys.
We will now quantify the size of this boost in the effective volume from multiple tracers with a large-scale clustering forecast.

\subsection{Large-scale clustering constraints under RT model uncertainty \label{subsec:applications_Fisher_gmu2}}

In this Section, we show that the lack of knowledge of the form and amplitude of the effect of RT on clustering can significantly degrade cosmological inference.
We make a straightforward Fisher forecast for the amplitude of the simplest possible class of RT effects - a dependence on $ \sum_n g_{2n} \mu^{2n}$ with unknown amplitudes $g_n$. 
For two galaxy populations $\delta_g^{(1)}$ and $\delta_g^{(2)}$, where $\delta_g^{(2)}$ is assumed to have \textit{no} explicit dependence on the line-of-sight fields (i.e.\ not selected by Ly$\alpha$ line emission), we can 
use the above expressions to quantify what survey configuration would be required to make a significant detection of this effect.
This is especially relevant for pilot LAE samples that can be assessed for the presence of such an effect before larger LAE surveys are undertaken (e.g.\ by Spec-S5).
Here we will work at the linear level for maximal clarity, though nonlinear line-of-sight dependent operators would be relevant on smaller scales \cite{Desjacques2018}.

For a zero mean Gaussian, such as $\delta_g^{(i)}$, the Fisher matrix for parameters $\alpha,\beta$ is
\begin{equation}
    F_{\alpha\beta} = \sum_{a = 1}^{N_{z}} \int_{0}^{1} d\mu \int_{k_{\mathrm{min}}}^{k_{\mathrm{max}}} N_{k,z_a} \mathrm{Tr}\left[C_{,\alpha}C^{-1} C_{,\beta} C^{-1}\right],
    \label{eqn:Fisher} 
\end{equation}
where $C_{ij}=\langle\delta_g^{(i)}\delta_g^{(j)}\rangle$ is the matrix of auto- and cross-spectra as functions of $(k,\mu)$, a subscript comma indicates a derivative.  The number of modes in a redshift shell is
\begin{equation}
    N(k,z_a) = \frac{k^2\,dk}{4\pi^2}\ V_{\rm{shell}}(z_a)
    \quad .
\end{equation}

The Fisher information for a single ($k,\mu$) mode on $g$ is
\begin{widetext}
\begin{equation}
   F_{gg}(k,\mu) = \frac{\mu ^4 n_2 \left[\tilde{\alpha}^2 n_2 (n_1
   P_g^{(1)}+2)+n_1 (n_1 P_g^{(1)}+1)\right]}{\left(b_1+f \mu
   ^2\right)^2 \left(n_1+\tilde{\alpha}^2 n_2\right)^2}
   \ R^2(k,\mu,z)
\label{eqn:Fisher_g} 
\end{equation}
\end{widetext}
where 
\begin{equation}
    \tilde{\alpha} = \frac{b^{(2)}+g\mu^2}{b^{(1)}+f\mu^2},
\end{equation}
and the fiducial case is given by $g=f$. 
Here we have neglected cross-stochasticity, which leads to a similar but longer expression.

We consider two survey configurations, one relevant for the future Spec-S5 survey, with an angular sky coverage of $11000~\mathrm{deg}^{2}$ with tracer sky densities of $N_\mathrm{LBG} = 2000~\mathrm{deg}^{-2}$, $N_\mathrm{LAE}=3000~\mathrm{deg}^{-2}$ (see Ref.~\cite{Beseuner25}, Table 2), and a more modest ``pilot'' configuration with $100~\mathrm{deg}^{2}$ sky area, and one third of the sky number densities.
For the pilot survey, we use $z_\mathrm{min}=2.5$, which reflects the difficulty of observing lower redshift Lyman-break galaxies (LBGs)\footnote{LBGs are not expected to have strong RT effects, as they are selected via a break with a range of wavelengths rather than the Lyman-$\alpha$ line. While it is indeed possible (and is not uncommon) for galaxies selected by their Lyman break to also be "LAEs" in the sense that they have Lyman-alpha emission in their spectra, the contribution of the LAE flux to the overall flux in a broad band is partially (or totally) washed out.} with DESI, but for Spec-S5 we use  $z_\mathrm{min}=2.1$, following Ref.~\cite{Beseuner25}.
The exact values of these 3 quantities (sky area, LAE sky density, and minimum redshift), for surveys that will be performed are at this point somewhat uncertain - we therefore explore their variation.

\begin{figure}
    \centering
    \includegraphics[width=0.5\textwidth]{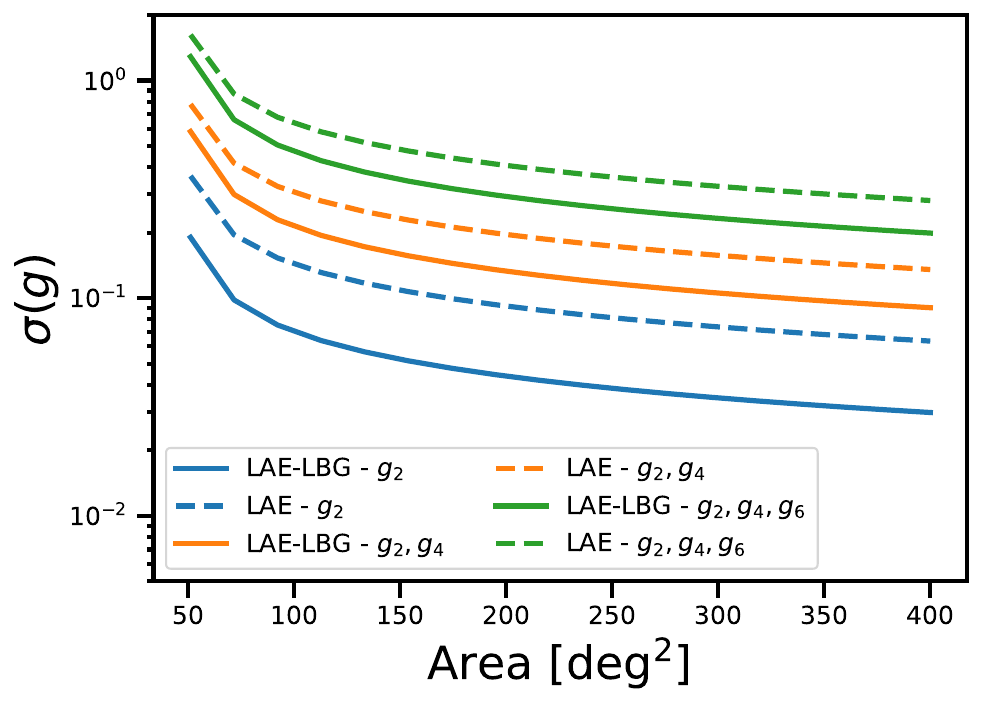}
    \caption{Marginalized (over linear bias, $\sigma_8$ and $g_{n>2}$) forecast on the simplest radiative transfer effect amplitude $g=g_2$ as well as more complicated RT models ($g_4$, $g_6$) in LAE single-tracer (dashed) and LAE+LBG multi-tracer forecasts (solid) for a small pilot survey.  The error on $g$ is plotted against survey area, with other assumptions described in the text.
    }
    \label{fig:area_plot}
\end{figure}

\begin{figure}
    \centering
    \includegraphics[width=0.5\textwidth]{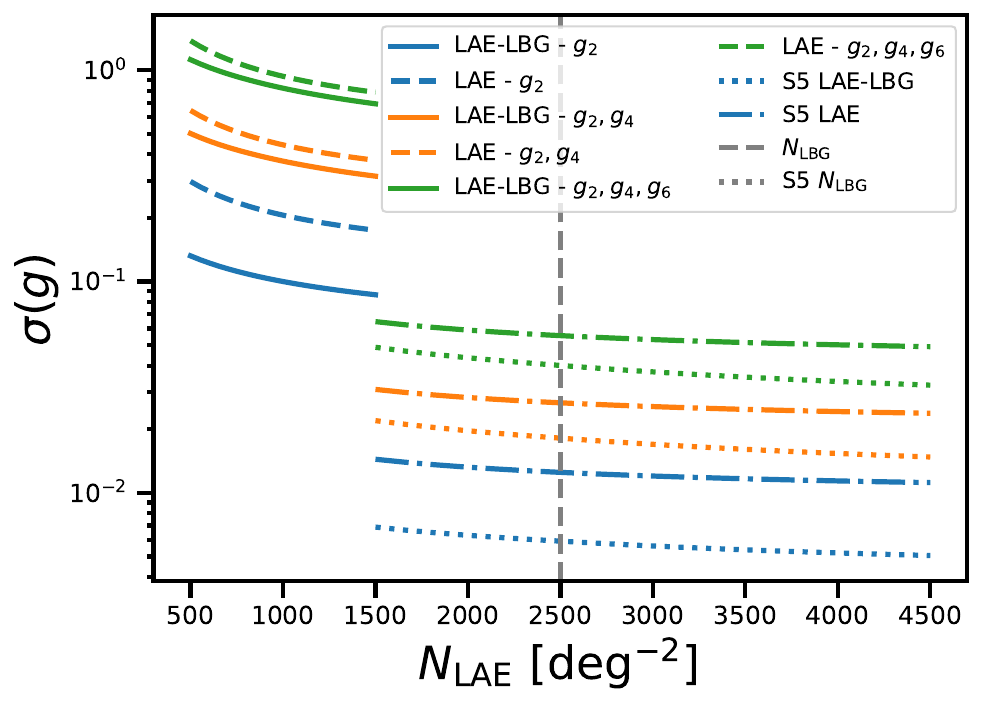}
    \caption{Similar to Fig.~\ref{fig:area_plot}, but as a function of angular sky number density $N_\mathrm{LAE}$.
    Curves at lower (higher) sky density show the multi-tracer forecasts for the pilot (S5) setup while blue, orange, and green curves account for increased freedom in the RT $\mu^{2n}$ model.
    Vertical dashed (dotted) gray lines indicate the LBG angular sky number density we assumed for the pilot (S5) survey.
    }
    \label{fig:ndens_plot}
\end{figure}

\begin{figure}
    \centering
    \includegraphics[width=0.5\textwidth]{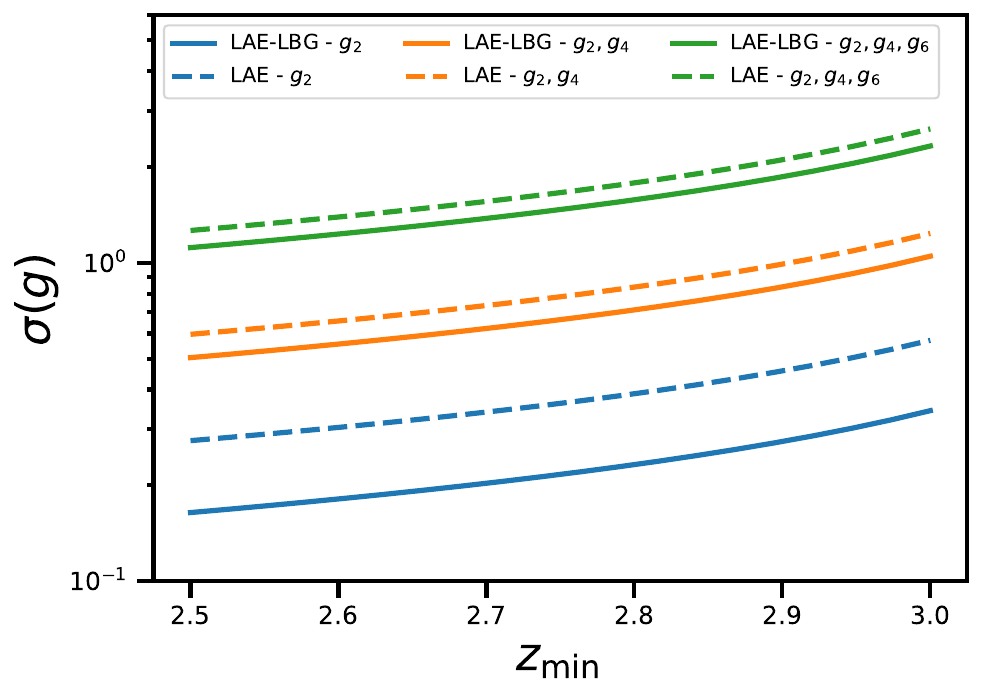}
    \caption{Similar to Fig.~\ref{fig:area_plot}, but as a function of minimum redshift $z_\mathrm{min}$ used for the forecast.
    The solid curves show the multi-tracer (LAE+LBG) result while the dashed curves show the LAE-only single-tracer result.
    Increasingly complex RT models of the form $\mu^{2n}$ are shown in blue, orange, and green; and, as expected, are accompanied with a loss of precision.
    All curves use the fiducial pilot area ($100~\rm{deg}^2$).
    }
    \label{fig:zmin_plot}
\end{figure}

We produce a set of numerical forecasts on the uncertainty of $g$ in this simplified context using the setup for LBGs and LAEs described above.
We use a $k_{\mathrm{min}}$ is determined by the fundamental mode of the survey configuration and 
$k_{\mathrm{max}}=0.1$, as appropriate for linear theory at these high redshifts.
We work with a minimum redshift of $z=2.5$ due to the technical difficulty of obtaining spectra for LBGs below this redshift.
For the LAE and LBG linear biases, we use  $b_{\mathrm{LAE}}=2.0 \frac{D(z=3)}{D(z)}$ \cite{2024:White_odin} and $b_{\mathrm{LBG}}=3.3 \frac{D(z=3)}{D(z)}$, respectively 
 \cite{Ruhlmann-Kleider:2024_lbg}.
We generally expect that the cross stochasticity for LAEs and LBGs is very small due to the different host halos of these populations.
Indeed, in Appendix~\ref{sec:perr}, we show that the cross noise is more than an order of magnitude smaller than the shot noise of the LAEs (which is smaller than that of LBGs) when considering the simulated galaxy samples of Ref.~\cite{Sullivan:2025_mtng_astrid_highz}.
For this reason we set $\bar{n}_{12}=-10\bar{n}_{\rm{LAE}}$ \footnote{Note that the cross-noise is negative, though this is largely irrelevant for our results due to its small amplitude.}, though once the cross noise is a factor of $\sim2-3$ smaller than the lower value of shot noise of the two tracers, results are largely insensitive to the size of the cross noise, and it can effectively be taken to zero.
We assume a fiducial Planck 2015 cosmology \cite{Planck15} throughout. 

Figure~\ref{fig:area_plot} illustrates the importance of knowing the complexity of the RT model as well as the benefit obtained from using multiple tracers for determining $g$.
In the absence of clear a priori information about RT complexity from theory, we consider the constraining power on the simplest RT term $g~ \mu^2$ in an increasingly complex RT model.
We express complexity using higher $\mu^{2n}$ moments - a $\mu^2$ term of the usual Kaiser form with $g_2=g\neq f$, a $g_4 ~\mu^4$ term, and $g_6 ~ \mu^6$ term.
Adding complexity to the model, expectedly, leads to significant degradation in the precision with which we can determine $g$ (and other large-scale amplitude parameters, like $f\sigma_8$).

The use of multiple tracers mitigates 
loss of constraining power - e.g., the $g_2,g_4$ model with LAEs and LBGs allows $g$ to be constrained at approximately the same precision as if considering only LAE data in a $g=g_2$ model (as seen from the near overlap of the blue dashed and solid orange curves).
This can be thought of as the boost in effective volume described earlier in Section~\ref{sec:orientation}.
We find similar gains from using multiple tracers over one tracer when the RT model is complex when varying the number density of LAEs (Figure~\ref{fig:ndens_plot}) and the minimum redshift of the survey (Figure~\ref{fig:zmin_plot}), for both the pilot and Spec-S5 (LBG) number densities.
Having established the potential impact of RT with unknown amplitude, we now turn to the case where both the RT amplitude and RT model (functional form) are unknown.

\section{Conditional density field \label{sec:cond_dens}}

\begin{figure*}[!htb]
    \centering
    \includegraphics[width=0.95\textwidth]{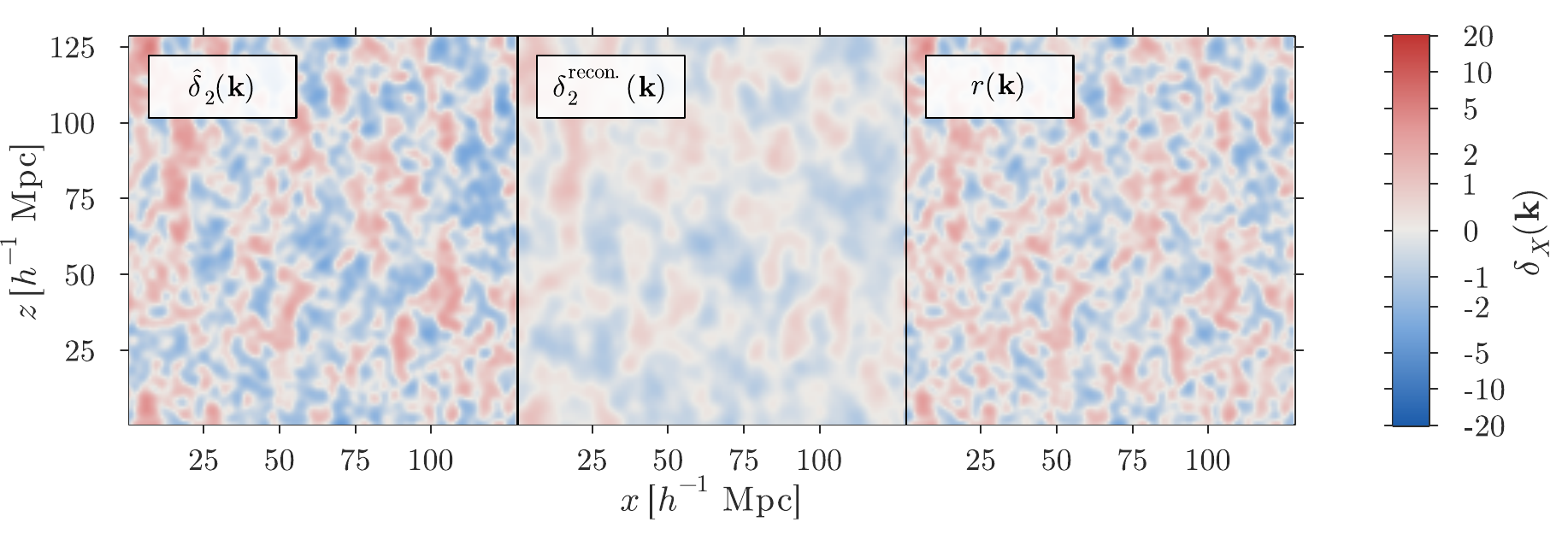}
    \caption{
    \textit{Model misspecification:} By exploiting the phase information contained in the LBG overdensity field, we are able to use sample variance cancellation to assess whether the RT model is misspecified.
    \textit{Left}: The true LAE density field $\hat{\delta}_2$
    \textit{Center}: Mean-field linear reconstruction $\delta_2^{\mathrm{recon.}}$ of the LAE tracer density using RT-free (LBG) tracer density $\hat{\delta}_1$, under an RT-misspecified ($g=f$) model for the LAEs. 
    \textit{Right}: Residual field between the reconstructed mean and the true LAE field. 
    Note the line-of-sight differences.
    Here we show transversely-averaged $20~\mathrm{Mpc}/h$ slices of each field and smoothed with a $2~\mathrm{Mpc}/h$ Gaussian filter, where the vertical axis corresponds to that of the line of sight. 
    }
    \label{fig:model_misspec_density}
\end{figure*}

\begin{figure}[h!]
    \centering
    \includegraphics[width=0.49\textwidth]{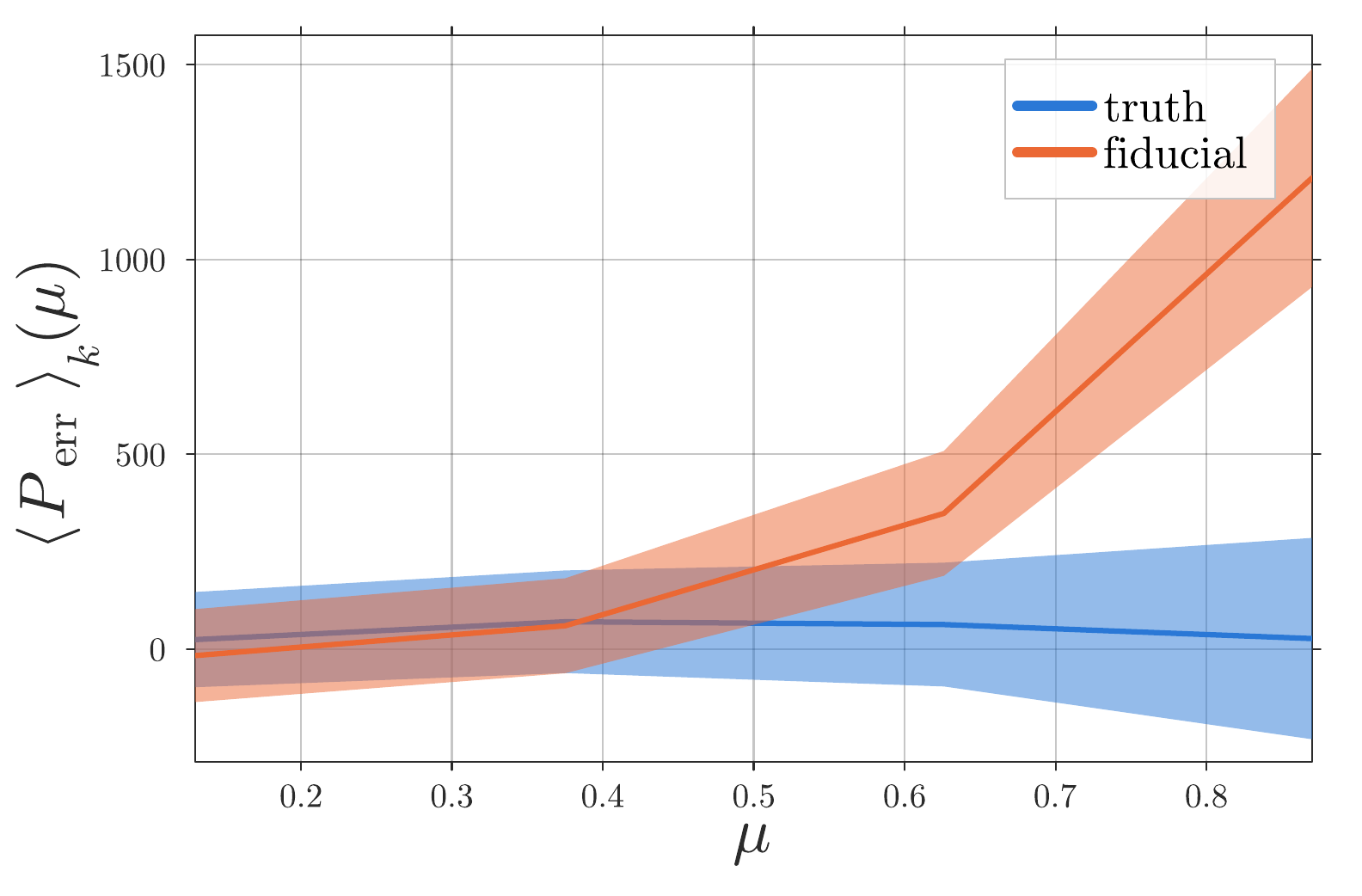}
    \caption{
    \textit{Model misspecification:}
    $\mu$-dependence of the wavenumber $k$-averaged power spectrum in 4 $\mu$ bins.
    We show both the error power spectrum from the linear reconstruction under a misspecified (trivial) assumed value of $g=f$ (orange) as well as the same reconstruction under the correct assumed value of $g=5f$ (blue).
    At low line-of-sight angle $\mu$, the error power spectrum of the two reconstructions is indistinguishable (as expected), while at high $\mu$ the difference is easily seen to be significant, indicating RT misspecification.
    Uncertainties are computed numerically using 100 draws from the conditional normal distribution in eqn.~\ref{eqn:d1_cond_d2}.
    }
    \label{fig:model_misspec_perrmu}
\end{figure}

To roughly quantify the extent to which we have failed to model RT effects, we now present a simple test of model misspecification. 
Here $\delta_g^{(i)}$ is the overdensity for galaxy sample $i$ and $\hat{P}^{(i)}$ is its \textit{observed} power spectrum (we will use $\hat{P}^{(i),(ij)}$ to refer to observed power spectra and  $P_g^{\{(i),(ij)\}} = P_g^{\{(i),(ij)\}}(\theta)$ to refer to theoretical power spectra.

For a fixed single mode, given fixed observed $\dtwo$, in the Gaussian case
\begin{equation}
    \delta_g^{(2)}(\vk) = \frac{\mathcal{K}^{(2)}(\mu)}{\mathcal{K}^{(1)}(\mu)}\done(\vk) + \left(\epsilon^{(2)}(\vk) - \frac{\mathcal{K}^{(2)}(\mu)}{\mathcal{K}^{(1)}(\mu)}\epsilon^{(1)}(\vk)\right)
\end{equation}
where $\epsilon^{(1)},\epsilon^{(2)}$ are (possibly dependent) Poisson distributed random fields and we have defined the effective $\mu-$dependent bias factors $\mathcal{K}^{(i)}(\mu) \equiv b^{(i)} + h \mu^2$ (where $h\in\{f,g\}$).
Here, $\done$ plays the role of the LBG field with standard growth rate $f$, while $\dtwo$ stands in for the LAE field with unknown parameter $g$ (or possibly more complicated real RT dependence).
Given observations of $\dtwo$, $\done$ we can attempt to estimate the distribution $P(\delta_g^{(2)}|\done)$ conditioned on the observed field $\done$. 
If the observed $\dtwo$ differs significantly from the predicted $\delta_g^{(2)}$, we can be confident that we have detected a deviation from our fiducial model.
This search for model misspecification will, of course, be limited by the noise impacting the observed fields $\done,\dtwo$. 
We assume the stochastic field can be approximated as Gaussian, as is typical on large scales for reasonably large sample number densities.

The distribution of $\delta_g^{(2)}$ conditioned on an observed realization of $\done$ is 
\begin{align}
    \label{eqn:d1_cond_d2}
    \delta_g^{(2)} &\sim \mathcal{N}(C_{12}C_{11}^{-1}\done ~, ~C_{22}-C_{12}C_{11}^{-1}C_{12}),\\
    &= \mathcal{N}\left(\left[\hat{P}_{12}/\hat{P}_{11}\right] \done ~, ~\hat{P}_{22}-\left[ \hat{P}_{12}^2/\hat{P}_{11}\right] \right),\nonumber
\end{align}
where the second line uses the fact we work on large scales to write the block covariances as diagonal.
Given a fiducial model for the covariance of the two fields, one can simulate an ensemble of realizations of $\delta_g^{(2)}$ and compare the resulting sampling distributions (even, in principle, $\vk$ mode by $\vk$ mode) to the observed field $\dtwo$, quantifying its deviation from the fiducial model.
A large significance of deviation indicates possible model misspecification, which can be followed up in further investigation to determine the source of the misspecification.
We can clearly see the sample variance cancellation in the context of eqn.~\ref{eqn:d1_cond_d2}, as, in the limit where the stochastic noise approaches zero, the conditional mean approaches the field times the relative bias and the conditional covariance approaches zero. 
In the sample-variance limit, we then recover the simple result that we can rescale the observed $\done$ by the effective relative bias to obtain $\delta_g^{(2)}$.
In practice, we find it most practical to use the best-fitting power spectra $P^{(ij)}(k,\mu)$ instead of the (noisy) data-estimated power spectra $\hat{P}^{(ij)}(k,\mu)$ for $P^{(1)}(k,\mu)$ and $P^{(12)}(k,\mu)$ and will do so for the practical demonstration of this Section below.
We note that while we apply this procedure in an extremely simple context, it is applicable even when the covariance is more realistic (e.g. due to the window function), so long as the situation remains approximately Gaussian.

Figure~\ref{fig:model_misspec_density} shows the conditional density procedure in action.
Working in a $250\,h^{-1}\mathrm{Mpc}$ periodic box, we adopt $80\%$ of the S5 survey number densities and linear bias model described in Section~\ref{sec:orientation} with $b^{(1)}=3.5$, $b^{(2)}=2.0$ at $z=3$ for the purposes of this demonstration.
For illustration, we have dramatically increased $g$ such that vertical (up-and-down) streaking due to line-of-sight clustering enhancement is readily visible\footnote{If $g$ were very close to $f$ for very noisy data, it might be advisable for the hypothetical observer to jump directly to Section~\ref{sec:opt_filt}.}.
The left panel shows the true LAE field while the center shows the reconstruction under the assumption that $g=f$.
It is immediately apparent that this is a poor reconstruction, as the very strong clustering along the line of sight is effectively absent, and as a result the residual field also shows these line-of-sight features (right panel), although the transverse and isotropic structure is decently captured.

Figure~\ref{fig:model_misspec_perrmu} makes this comparison slightly more quantitative, showing that the error power spectrum (the power of the residual field between the conditional density reconstruction and the true LAE field) exhibits strong $\mu$ dependence at higher $\mu$ values.
Specifically, these are the shot-noise-subtracted power spectra, where the error power spectra in the noise dominated limit, $P_\mathrm{err.,~n.d.}$, take the form
\begin{equation}
    P_\mathrm{err.,~n.d.}(k,\mu) = \frac{1}{\bar{n}_2} - 2 \frac{\mathcal{K}^{(2)}(k,\mu)}{\mathcal{K}^{(1)}(k,\mu)}\frac{1}{\bar{n}_{12}} + \left( \frac{\mathcal{K}^{(2)}(k,\mu)}{\mathcal{K}^{(1)}(k,\mu)}\right)^2 \frac{1}{\bar{n}_{1}},
    \label{eqn:err_power_noise_domin}
\end{equation}
where the factor $\mathcal{K}^{(2)}$ carries the $g$ dependence.
After subtracting this term, the true RT model is consistent with an error power spectrum of zero (within the errorbar implied by the Poisson stochasticity of $\done$) and the incorrect fiducial case gives a significant error power spectrum at high $\mu$.

Presented with this potential clear signal of strong unmodeled angular dependence, we now turn to a model-agnostic method for determining the functional form of this dependence in the next Section.

\section{Optimal filtering  \label{sec:opt_filt}}

\begin{figure*}[htb!]
    \centering
    \includegraphics[width=0.49\textwidth]{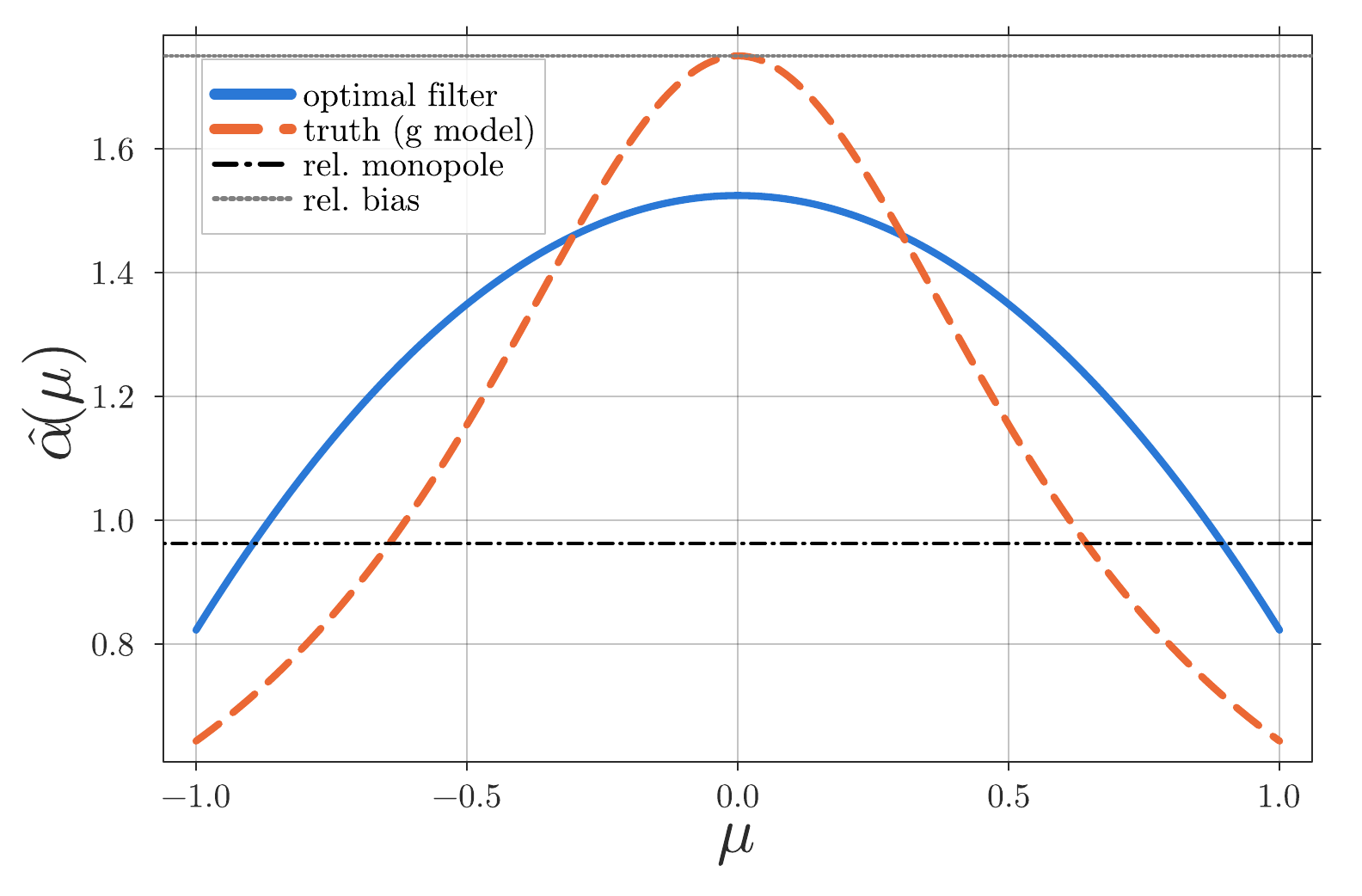}
    \includegraphics[width=0.49\textwidth]{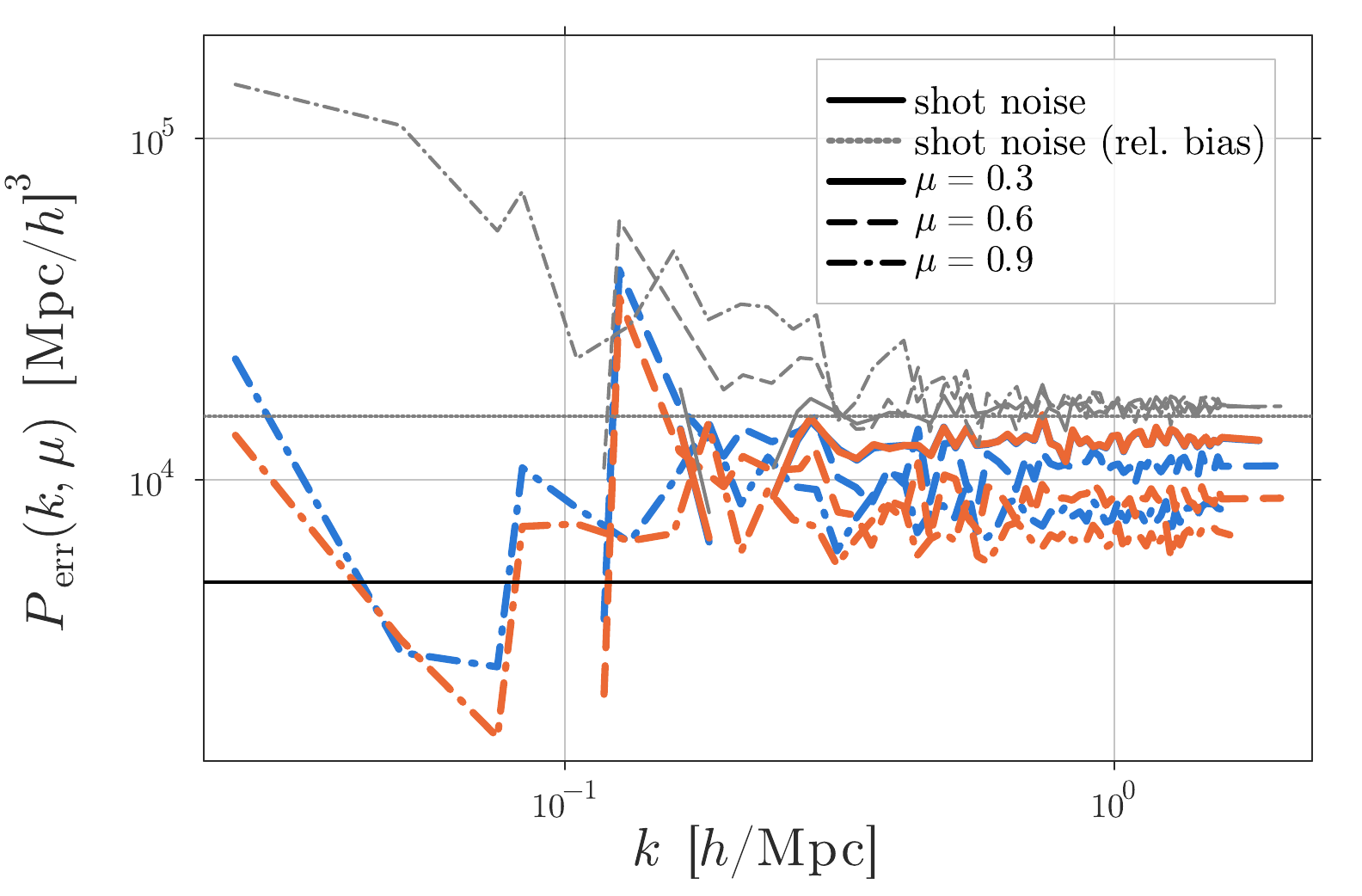}
    \caption{
    \textit{Optimal filtering:}
    \textit{Left:} 
    The estimated optimal filter $\alpha$ for a linear mock survey of LBGs and LAEs.
    This filter (blue solid) allows us to determine the functional form of the RT effect optimally (in a mean-squared sense).
    We also show the true relative $\mu$-dependence between the two fields, which is exact in our linear toy model (orange dashed).
    In addition, we show the monopole prefactor in eqn.~\ref{eqn:c_0_simpl} (dash-dotted black) and the bias ratio $b_1/b_2$ (gray dotted) as horizontal lines. 
    The best quadratic approximant given by the optimal filter largely captures the $\mu$-dependence of the RT model.
    \textit{Right:} The error power spectrum $P_\mathrm{err}(k,\mu)$ in several $\mu$ bins (solid->dashed->dash-dotted line styles are higher $\mu$) for both the optimal filter (blue), the truth (red), and the relative bias (gray).
    Missing points in a given $\mu$ bin are due to a lack of modes on the discrete grid.
    For reference, we also show the shot noise expectation for the reconstructed field, which sets a lower bound on our reconstruction noise.
    The optimal filter improves significantly with respect to the simple relative bias approximation baseline.
    }
    \label{fig:optimal_filtering}
\end{figure*}

We now describe a more flexible procedure to search for possible $\mu-$dependent systematics in the galaxy field when multiple tracers are available.
Our focus here will be on angular ($\mu$-dependent) systematics, such as the possible RT effects in the LAE galaxy field.
In the hypothetical scenario that the conditional density diagnostic test of Section~\ref{sec:cond_dens} has indicated the presence of unmodeled RT effects, or if one believes otherwise that they should be present, the following procedure provides a strategy for optimally obtaining an estimate of the functional form of the $\mu$ dependence.

Using the same notation for power spectra as in Section~\ref{sec:cond_dens} now in a given redshift bin and k bin\footnote{We are neglecting the impact of window functions here, since the mapping from $P_\ell^{\{(i),(ij)\}}$ to $\hat{P}_\ell^{\{(i),(ij)\}}$ will be the same for all auto- and cross-spectra if we match the footprints and $dN/dz$.}, we consider the power spectrum multipoles of two observed galaxy fields 
$\hat{\delta}_g^{(1)}(\mu) = \hat{\delta}_g^{(1)}(\mu,k_i,z_j)$, $\hat{\delta}_g^{(2)}(\mu) = \hat{\delta}_g^{(2)}(\mu,k_i,z_j)$ 
at a fixed redshift $z_j$ and in a particular wavenumber bin, $k_i$.
We decompose the generic $\vk$ dependence into $(k,\mu)$ via the usual assumed line-of-sight in the plane-parallel approximation, and work at linear order in the density field.

The residual between our two galaxy fields is
\begin{equation}
    \label{eqn:resid_def}
    r(\mu) = \done(\mu)-\alpha(\mu)\dtwo(\mu).
\end{equation}
Ideally we would find a function $\alpha(\mu)$ such that $r(\mu)=0$.  
However, in the presence of noise this will not be possible; though we can pursue a weighted residuals\footnote{This is closely related to the method of weighted residuals which is used, for instance, in variants of the Finite Element Method.} approach by introducing a weighting function $w(\mu)$ and minimizing the weighted residual to obtain the optimal $\hat{\alpha}$ approximating $\alpha$
\begin{align}
    \label{eqn:weight_residual}
    \min_{\hat{\alpha}}\langle w(\mu)r(\mu)\rangle = 0.
\end{align}
Since we are most interested in $\mu-$dependent RT effects, we express $\hat{\alpha}$ in a basis of Legendre polynomials
\begin{widetext}
\begin{equation}
    \label{eqn:alpha_approx}
    \hat{\alpha}(\mu) = \sum_{\ell=0}^{\ell_{\mathrm{max}}}c_{\ell} \mathcal{L}_\ell(\mu).
\end{equation}
\end{widetext}
The minimization problem for determining the optimal $\hat{\alpha}$ then reduces to an optimization over the $c_\ell$ coefficients;
\begin{align}
    \label{eqn:weight_residual_2}
    \min_{c_\ell}\langle w(\mu)r(\mu)\rangle = 0.
\end{align}

What is the metric by which we want to determine the $c_\ell$ coefficients?
Different metrics will lead to different senses of optimality; here we select the optimal approximation in the least-squares sense, as this is also the approximation that minimizes the metric of most interest here, the error power spectrum $P_{\mathrm{err}}(k,\mu) = \langle r^\star(k,\mu)r(k',\mu')\rangle$. 
In this case, the weighting function is, for arbitrary coefficients $a_\ell$ (that drop out of the final expression),
\begin{align}
    \label{eqn:lsq_weight}
    w(\mu) &= \sum_{\ell=0}^{\ell_\mathrm{max}} a_\ell w_\ell(\mu)\\
    &= \sum_{\ell=0}^{\ell_\mathrm{max}}a_\ell\frac{\partial r(\mu)}{\partial c_\ell}, \nonumber
\end{align}
which minimizes $\langle r^\star r\rangle$. 

The simultaneous expansion of $\hat{\alpha}(\mu)$ and the weight function $w(\mu)$ in Legendre polynomials allows us to find the values of $c_\ell$ by solving a linear system
\begin{equation}
    \label{eqn:lin_sys}
    L\mathbf{c} = \mathbf{v}
\end{equation}
where
\begin{widetext}
\begin{equation}
L = \begin{pmatrix}
\langle \dtwo(\mu)\dtwo(\mu)\mathcal{L}_0(\mu)\mathcal{L}_0(\mu)\rangle ~~...~\langle \dtwo(\mu)\dtwo(\mu)\mathcal{L}_0(\mu)\mathcal{L}_{\ell_{\mathrm{max}}}(\mu)\rangle\\
\vdots ~~~~~~~~~~~~~~~~~~~~~~~~~~~~~~~~~~~~~~~~~~~~~~~~~~~~~~~~~~ \vdots\\
\langle \dtwo(\mu)\dtwo(\mu)\mathcal{L}_{\ell_{\mathrm{max}}}(\mu)\mathcal{L}_0(\mu)\rangle~~...~\langle \dtwo(\mu)\dtwo(\mu)\mathcal{L}_{\ell_{\mathrm{max}}}(\mu)\mathcal{L}_{\ell_{\mathrm{max}}}(\mu)\rangle
\end{pmatrix},
\label{eqn:L_def}
\end{equation}
\end{widetext}
is a symmetric matrix, and
\begin{equation}
\mathbf{c} = \begin{pmatrix}
c_{0}\\
\vdots\\
c_{\ell_{\mathrm{max}}} \\
\end{pmatrix}
\mathrm{\ and\ }
\mathbf{v} = \begin{pmatrix}
\langle \done(\mu)\dtwo(\mu)\mathcal{L}_0(\mu)\rangle \\
\vdots\\
\langle \done(\mu)\dtwo(\mu)\mathcal{L}_{\ell_{\mathrm{max}}}(\mu)\rangle \\
\end{pmatrix} \quad,
\end{equation}
are vectors.
This has the expected structure of a Wiener filter.

For the Kaiser model we would have $\ell_{\mathrm{max}}=4$ and all of the odd multipoles zero, and, in the large-scale regime we work in, the hexadecapole is significantly smaller than the monopole and quadrupole.
If we fix $\ell_\mathrm{max}=2$ \footnote{It is not enough to simply neglect $P^{(i)}_{\ell\ge 4}$ as, with $\ell_\mathrm{max}=4$, terms will arise from, e.g., the non-zero $\mu^1 \mu^3$ terms from $P_{22}$ in eqn.~\ref{eqn:L_def}.} the resulting optimal coefficients are
\begin{equation}
    \label{eqn:drophex_c_ells}
    \begin{pmatrix}
        c_0\\
        c_2
    \end{pmatrix}
    = N
    \begin{pmatrix}
        \hat{P}_0^{(12)}\hat{P}_0^{(2)} + (10/7)\hat{P}_0^{(12)}\hat{P}_2^{(2)} - 5 \hat{P}_2^{(12)}\hat{P}_2^{(2)} \\
        \vphantom{0} \\
        5[\hat{P}_2^{(12)}\hat{P}_0^{(2)} - \hat{P}_0^{(12)}\hat{P}_2^{(2)} ]
    \end{pmatrix}
\end{equation}
where $N^{-1}=(\hat{P}_0^{(2)})^2 + (10/7)\hat{P}_0^{(2)}\hat{P}_2^{(2)} - 5 (\hat{P}_2^{(2)})^2$.

Note that in the limit that $\hat{P}_2\to 0$ one obtains the well-known result: $c_0=\hat{P}^{(12)}_0/\hat{P}^{(2)}_0$.
In the limit where the noise is small, this is just the relative effective bias of the two tracers.
More generally, when neglecting cross-stochasticity, we recognize a simple rescaling of the single tracer Wiener filter (eqn.~\ref{eqn:s_sn_1}), 
\begin{equation}
\label{eqn:c_0_simpl}
c_0\stackrel{\bar{n}_{12}\to 0}{=} \sqrt{\frac{(\mathcal{K}^{(1)}\mathcal{K}^{(2)})_0}{(\mathcal{K}^{(2)})^2_0}}\left(\frac{\bar{n}_2  P_{g,0}^{(2)}}{1+\bar{n}_2P_{g,0}^{(2)}}\right),
\end{equation}
where $P_{\ell}^{(ij)} = P_{g,\ell}^{(ij)} + N^{(ij)}$.
At low $k$, where $\bar{n}P\gg 1$, the $c_\ell$ become $k$-independent.  When $\bar{n}P\ll 1$ the fields are shot-noise dominated and no cancellation is possible: $c_\ell\approx 0$.
If the data are noisy, we recommend using the best-fit theoretical model for computing $c_\ell$ to avoid dividing by noisy quantities\footnote{The assumption of a theoretical model will introduce an error in $c_\ell$, which can be mitigated by adding more data.
However, this error should be small for RT models that are not obviously misspecified.}.

Figure~\ref{fig:optimal_filtering} shows the optimal filter for the linear LAE-LBG mock configuration described in Section~\ref{sec:cond_dens}, but for the pilot number densities.
In the left panel, we see what is effectively the relative angular transfer function between the two fields $\done$ and $\dtwo$.
In the simple Kaiser-style model for RT with unknown $g$, we show (dashed orange) the true filter that should result in minimal mean residual (up to the dominant stochasticity).
This dependence is complex as it involves the ratio of the two Kaiser polynomials in $\mu$ for $\done,\dtwo$, and it is not easily captured by a quadratic function.
However, the optimal filter (in the MSE sense, solid blue) following eqn.~\ref{eqn:drophex_c_ells} is indeed the best possible quadratic function approximating the angular dependence of the relative transfer function at $\ell_\mathrm{max}=2$.
For comparison, we also show the case of $\ell_{\rm max}=0$ of eqn~\ref{eqn:c_0_simpl} (when $\bar{n}_2 P^{(2)}_{g,0}$ is large, black dash-dotted) and the bias ratio that is exact in the $\mu=0$ case (gray dashed).

The right panel of Figure~\ref{fig:optimal_filtering} quantitatively shows the performance of these various filters in terms of the error power spectrum $P_\mathrm{err}(k,\mu)$ in several $\mu$ bins (various curve shades).
Good performance is attained when the error power spectrum of the residual of filtered field saturates the lower bound provided by the dominant shot noise (in this case, of the lower-number-density LBGs).
For reference we show the truth (red) and the LBG shot noise (dashed black).
Due to the presence of noise in the synthetic data realizations that are rescaled by the relative transfer function, even the true relative transfer function fails to furnish an error power that reaches the LBG shot noise.
The optimal filter (blue) overall does nearly as well as the true relative transfer function, suggesting that we gain important information from it, especially relative to the too-simple relative bias scaling (gray dashed).
However, a mild performance gap due to the beyond-quadratic nature of the true model, even in this simple Kaiser-type case, prevents the optimal filter from saturating the lower bound provided by the LBG shot noise. 
To obtain a more complete estimate of the amplitude of the angular dependence parameterized in the model favored by the optimal filter, we now turn to a more comprehensive tool.

\section{Quadratic estimator  \label{sec:qe}}

After performing optimal filtering, we expect that  
the functional form of the angular dependence of the field of interest is more-or-less known, albeit with some uncertainty due to noise in the data.
With the best-fit parameters of this fixed model, we then have a reasonable starting point from which to wring out as much information as possible in the determination of these parameters.
We accomplish this by way of a quadratic estimator (QE) for these parameters (e.g. $g_2, g_4,...$).

\subsection{Single tracer \label{subsec:qe_single_tracer}}
Let us first consider a quadratic estimator for $g=g_2$ from a single tracer (e.g. LAE) power spectrum $P^{(2)}$. 
For simplicity, this assumes we found no evidence for $g_4$, $g_6$, etc in Section~\ref{sec:opt_filt}, though the generalization to include $g_4,g_6$ case follows immediately.
The naive quadratic estimator for $g$ follows directly from the QE for the estimated power spectrum bandpowers, which we could derive as the maximum likelihood estimator for a Gaussian likelihood 
\cite[e.g.,][]{Tegmark:1998_galpower_qe,Hamilton1997:qe,Tegmark1997:qe_kl,BondJaffeKnox1998:qe,Philcox2021:qe,Seljak2017:recon}.
We work under the usual assumptions: that the fiducial parameter estimate is sufficiently close to the global optimum that the problem is effectively unimodal, that we are within the curvature scale of the quadratic likelihood surrogate, and that the covariance model functional form is not misspecified (following optimal filtering; though we will consider \textit{parameter} misspecification). 
The estimator is $\hat{g} = F_{gg}^{-1} q_{g}$ with
\begin{equation}
    \label{eqn:naive_qg}
    q_{g} \equiv \frac12 d^\dagger C^{-1} C_{,g} C^{-1} d - b_g,
\end{equation}
where the data covariance
is $C = S + N$.
Unless otherwise stated, all factors of $C$ are factors of fiducial covariance and unadorned factors of $g$ indicate the fiducial parameter value.
We take the usual choice of 
weighting function $E_g = C^{-1} C_{,g} C^{-1}$, which is the maximum likelihood weight when the fiducial covariance is the true covariance.
Explicitly, we have
\begin{align}
    &C_{,g} = 2(b^{(2)} \mu^2 + g \mu^4) P_m,\\
    &C_{,g g} = 2\mu^4 P_m,
\end{align}
and where
\begin{equation}
    b_g = \frac12 \mathrm{tr}[E_gN]
    \label{eqn:bg_naive_N}
\end{equation}
ensures the estimator is unbiased for a linear parameter.

We note that the galaxy field covariance from $P^{(2)}$ is quadratic in $g$ rather than linear:
\begin{equation}
    \label{eqn:ctrue_exp}
    C^{\mathrm{true}} \approx C + (\delta g) C_{,\delta g}+ \frac{(\delta g)^2}{2} C_{,\delta g \delta g},
\end{equation}
where $\delta g \equiv g^{\mathrm{true}}-g$,
such that $\delta C \equiv C^{\mathrm{true}} - C \approx \delta g C_{,\delta g}+ \frac{(\delta g)^2}{2} C_{,\delta g \delta g}$.
For our model where any unknown RT parameter enters quadratically in the covariance, this approximant is actually exact, and we will use this going forward. 
The nonlinearity of the covariance has two important, if easy, consequences: 1) we must modify the bias term $b_g$, which ends up being parameter-dependent, to avoid making errors of order $(\delta g)^2$ and, 2) there is an additional term contributing to suboptimality in the nonlinear case, which we may also correct for. 
It is also possible to forgo the quadratic estimator and just optimize the likelihood directly via exact (quasi-)Newton steps \cite[e.g.,][]{BondJaffeKnox1998:qe} - this is exactly the same as the iterated corrected quadratic estimator we will now describe.

\paragraph{Bias}
Unlike the case of a covariance that is linear in the parameters, the naive $q_g$ estimator is biased even after the usual correction $b_g$ provided in eqn.~\ref{eqn:bg_naive_N}.
The correct debiasing term can be derived from the maximum likelihood estimator when the true covariance is the fiducial covariance,
\begin{align}
    b_g &=  -g ~F_{gg} + \frac12 \mathrm{tr}[C^{-1}C_{,g}]\\
    &= \frac12 \mathrm{tr}[E_g (C-gC_{,g})], \nonumber\\
    &=\frac12 \mathrm{tr}[E_g (C-S)]= \frac12 \mathrm{tr}[E_g N], \nonumber
    \label{eqn:bg_lin}
\end{align}
where the last line assumes linearity,
and we would have
\begin{align}
    \langle \hat{g} \rangle = g-\frac12F_{gg}^{-1}\mathrm{tr}[C^{-1}C_{,g}]
\end{align}
where, for a linear model for parameter $\theta$, we have $S = \theta~C_{,\theta}$.
However, for the quadratic model, we instead have
$S = C + (\delta g)~C_{,\delta g} ~+ \frac12 (\delta g)^2~C_{,gg}$, and so must replace $C-S$ on the last line of eqn.~\ref{eqn:bg_lin} with
\begin{align}
    b_g &= \frac12 \mathrm{tr}\bigg[E_g\bigg(C+\frac{(\delta g)^2}{2}C_{,gg}\bigg)\bigg],\\
    &= \frac12 \mathrm{tr}[C^{-1}C_{,g}]+ \frac12 \mathrm{tr}\bigg[\frac{(\delta g)^2}{2}E_gC_{,gg}\bigg].
\end{align}
To correct for this factor, we must know $\delta g$, which of course we do not, since we do not know $g^{\mathrm{true}}$.
However, we can estimate it (up to corrections of $(\delta g)^2$) by the current value of the quadratic estimate $\delta \hat{g} \equiv \hat{g}-g$, which turns out to be an excellent estimate.

\begin{figure*}
    \centering
    \includegraphics[width=0.49\textwidth]{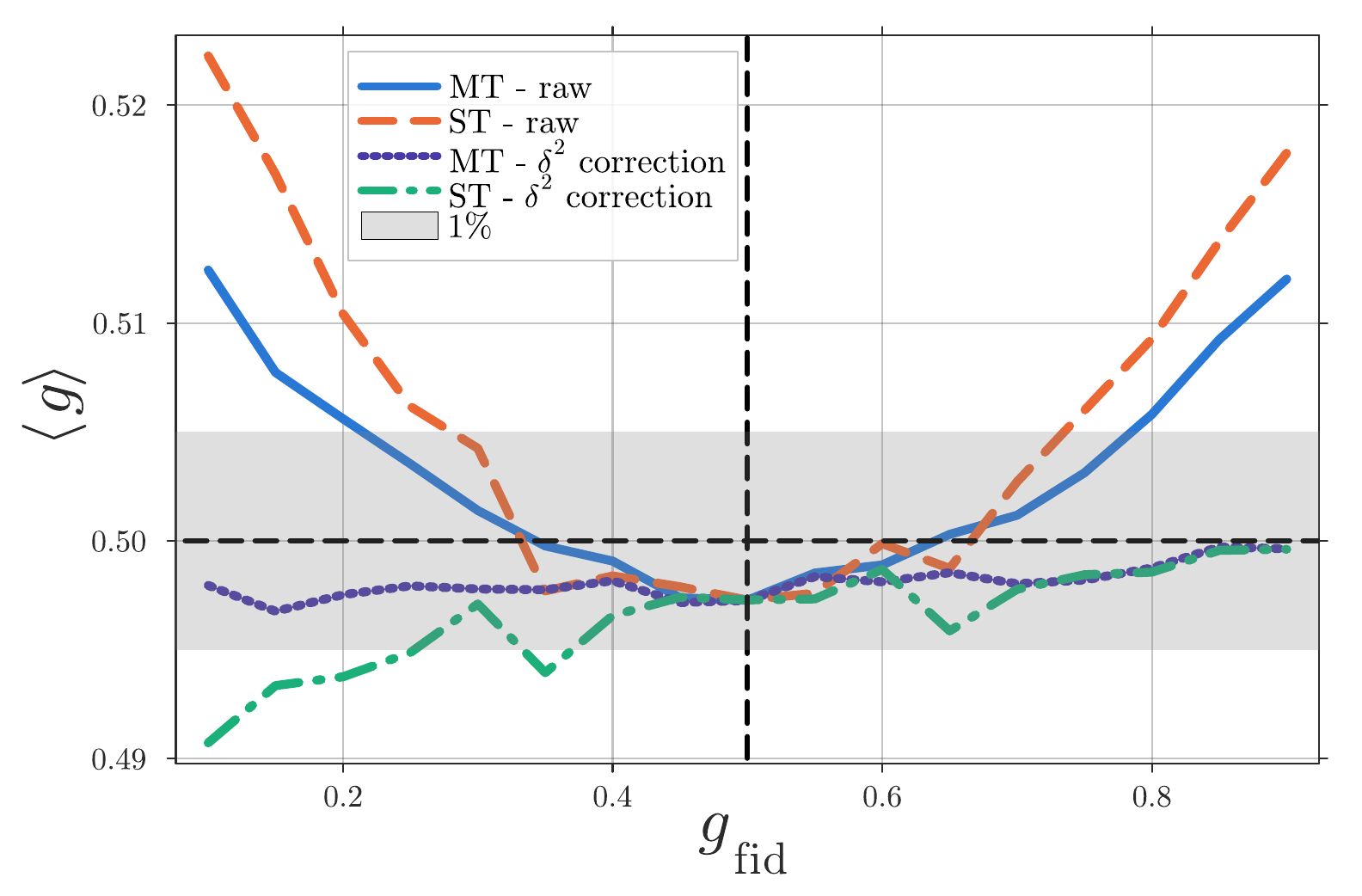}
    \includegraphics[width=0.49\textwidth]{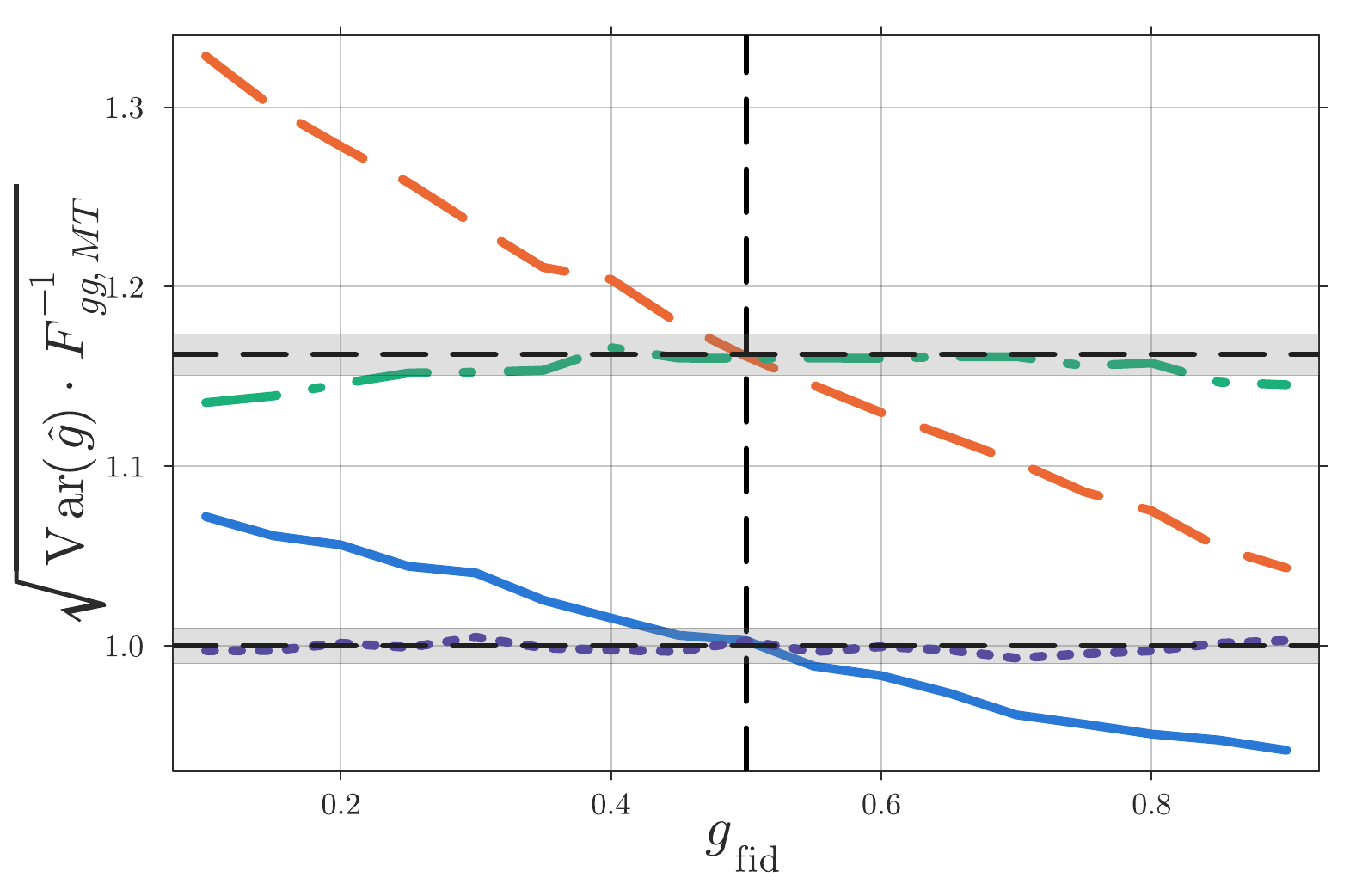}
    \caption{
    \textit{Quadratic estimator:}
    The corrected quadratic estimator for $g$ has lower and less biased uncertainty in the multi-tracer (MT) case than in the single-tracer (ST) case.
    \textit{Left}: We show inferred values of $g$ from the quadratic estimator as a function of $\delta g$, the difference of the fiducial value of $g$ from the true value used to generate the data.
    We show both the bias on the estimator from the LAEs alone (dashed orange) and from the LAEs in conjunction with the LBGs (solid blue), finding that applying the leading correction to both the MT (purple dotted) and ST (green dash-dotted) cases significantly reduces the bias, though the MT result is slightly more stable.
    \textit{Right:} The curves have the same meaning, but we show the estimated uncertainty on $\hat{g}$ rather than the estimated value of $g$ itself.
    The upper horizontal line gives the ratio of the ST Fisher matrix to that of the MT case for the true value of the parameter $0.5$.
    There we see that the multi-tracer case lowers the uncertainty by 20\% when considering $g$ alone, and demonstrates a smaller bias on the error after applying the leading correction to the variance than the single tracer case, which retains a $>1\%$ bias on the error with an apparent quadratic dependence on the fiducial value of $g$.
    }
    \label{fig:qe_deltag}
\end{figure*}

\paragraph{Variance}
As for all quadratic estimators, $q_g$ is also suboptimal unless the true covariance is exactly equal to the fiducial covariance $C^{\mathrm{true}}=C$ (where, again, we take $C=C^{\mathrm{fid}}$).
However, due to the nonlinearity of the covariance of $\hat{g}$, the degree to which optimality is violated is increased due to the presence of extra terms.
Specifically, carrying through the usual exercise of computing the variance of the QE as applied to a GRF (with only trivial trispectrum), the variance is \footnote{Here we compute only the variance in this simple demonstration to keep the discussion relatively index-free, but, of course, the covariance of the estimator of two quadratic parameters proceeds in exactly the same way.}
\begin{align}
    \label{eqn:var_qg}
    \mathrm{Var}(q_g) &= F_{gg}\\
    &\quad+ \delta g ~\mathrm{Tr}[(C^{-1}C_{,g})^{3}] \nonumber\\
    &\quad+ \frac{(\delta g)^2}{2} \bigg\{ \mathrm{Tr}[(C^{-1}C_{,g})^{4}] \nonumber\\
    &\quad\quad\quad\quad\quad+ \mathrm{Tr}[(C^{-1}C_{,g})^{2} (C^{-1}C_{,gg})] \bigg\} \nonumber\\
    &\quad+ \mathcal{O}((\delta g)^3)\nonumber.
\end{align}
From this expression we can see that there is (as usual for a linear parameter) both a linear and quadratic bias in $\delta g$, but
where the second quadratic-in-$\delta g$ term on the fourth line is unique to the case of the quadratic parameter dependence, as it is absent in the case of a linear dependence of the covariance on the parameter of interest.
In Section~\ref{subsec:qe_demo}, we will show that subtracting the leading $\mathcal{O}(\delta g)$ term
largely corrects the estimator's variance.

\subsection{Multiple tracers \label{subsec:qe_multitracer}}
Now let us generalize the previous discussion to the case of multiple tracers with data vector $\hat{\delta} \equiv \begin{pmatrix}
    \done\\
    \dtwo
\end{pmatrix}$ that we consider in this work.
In this case, we must only slightly generalize the quadratic estimator for $g$
\begin{equation}
    q^{(12)}_{\delta g} \equiv
    \hat{\delta}^\dagger
    C^{-1} C^{(12)}_{,\delta g}  C^{-1} \hat{\delta} - b^{(12)}(\delta g),
\end{equation}
where $C$ is now block diagonal,
\begin{equation}
    C = \begin{pmatrix}
        C^{(11)} & C^{(12)}\\
        C^{(12)}& C^{(22)}
    \end{pmatrix},
\end{equation}
with
$C^{(ij)} = S^{(ij)} + N^{(ij)}$, $N^{(11)}=\bar{n}_1^{-1}$, $N^{(12)}=\bar{n}_{12}^{-1}$, $N^{(22)}=\bar{n}_{2}^{-1}$,
and the derivative matrices are also block diagonal, 
where the new cross-covariance terms are
\begin{align}
    C^{(12)}_{,\delta g} &= 2(b^{(1)} \mu^2 + f\mu^4) P_m,\\
    C^{(12)}_{,\delta g \delta g} &= C^{(11)}_{,\delta g \delta g} =C^{(11)}_{,\delta g} = 0.
\end{align}
With these modifications, the bias correction $b^{(12)}$ takes the same form as before.
The addition of the second tracer adds no difficulties of the type encountered for the original $\hat{g}$ estimator, as $P_{12}$ is actually linear in $g$, so all the usual statements for linear covariance models (e.g., that apply to power spectrum bandpowers) carry over.
For sufficiently large signal-to-noise, we find that the multi-tracer case produces a $\mathcal{O}(\delta g^2)$ term that dominates the $\mathcal{O}(\delta g)$ error in the variance, but the linear correction of eqn.~\ref{eqn:var_qg} is enough to get within 1\% of the CRLB (see the right panel of Fig.~\ref{fig:qe_deltag}).

\subsection{Optimally recovering RT amplitude \label{subsec:qe_demo}}
We now apply this QE for $\delta g$ to a simple GRF mock scenario, similar to the one described in Sections~\ref{sec:cond_dens}, \ref{sec:opt_filt}, for the pilot number densities as a demonstration.
In this example, we consider the case where the true RT amplitude value is $g=0.5$ and apply the quadratic estimator at a series of fiducial $g_\mathrm{fid}$ values (ranging from $0.9-1.8$, as we might anticipate from particularly noisy data that suggests $g\gtrsim 1$). 
We use a $N_g=16^3$ grid ($k_\mathrm{max}=0.2~h~\mathrm{Mpc}^{-1}$) and discard $\mu=0$ modes, which provide no RT information, on the grid.
For this simple demonstration, we neglect realistic observational effects (such as due to the survey geometry) and uncertainties in the bias and cosmological parameters, so the absolute errorbars should be considered optimistic and are not to be interpreted at face value.
We simply aim to illustrate the relative improvement due to the presence of multiple tracers as well as the efficacy of the derived QE corrections. 

Fig.~\ref{fig:qe_deltag} shows the recovery of the value of $\delta g$ for an array of initial guesses for $g$ (as we might anticipate from particularly noisy data).
We show both the single-tracer estimates and multi-tracer estimates before and after applying the corrections above.
To assess unbiasedness of the estimator and its variance in the pilot-like survey configuration, we use 50,000 Monte Carlo realizations. 
In the left panel, we see that the single-tracer and multi-tracer results are similar, with a somewhat smaller bias for the multi-tracer case.
In terms of the optimality of the estimator, we see that both tracers have an error proportional to $\delta g $ that leads to a linear deviation in the variance from the CRLB, but that this deviation is removed at the 1\% level when applying the leading correction of eqn.~\ref{eqn:var_qg}.
We note that in several more favorable signal-to-noise regimes we explored, the single-tracer result looks qualitatively similar, but the multi-tracer result has as its dominant error the curvature-type correction at order $\delta g^2$.
In this simplistic scenario where only $g$ is marginalized, the multi-tracer QE provides a moderate, yet sizable, gain in the constraining power on $g$; though, as demonstrated in Section~\ref{sec:orientation}, the gain is much larger for marginalization over multiple large-scale ($\mu$-dependent or -independent) amplitudes as would occur in cosmological data analysis.

\section{Discussion \& Conclusion \label{sec:conc}}
Modern cosmology enjoys high-fidelity observations of multiple tracers of the density field over wide areas and a long redshift lever arm.
As the statistical power of this lever arm improves over the next decade with high-redshift observations, the need for systematics mitigation grows.
We have plotted a course to constrain systematics in galaxy survey data when observations of two tracers of the cosmic density field are available in the same volume and one of the tracers is likely to be free of the bias of interest. 
Assuming the role of a hypothetical observer faced with measurements of an LSS tracer field containing unknown angular dependence\footnote{Here we refer to unknown dependence of the Fourier-space LSS field on $\vk$ modes with different projections onto the observer's line of sight. We do not refer to the ``angular systematics'' that vary in the plane of the sky (e.g. ground-based seeing, stellar density) that may contaminate an observed LSS density field; though, of course, these must also be corrected for in a real data analysis.}, we offer up a galaxy-clustering enchiridion in 3 steps.
First, we suggest generating field-level realizations conditioned on the ``clean'' density field in order to assess the strength of any unmodeled angular dependence of the field of interest.
If this test suggests unmodeled systematics then the functional form of this dependence can be determined through the use of the optimal filter technique we derived.
Finally, after establishing the model class with reasonable certainty, we provide an optimal quadratic estimator that can be applied to maximally constrain the parameters of the initially unknown angular dependence.

To illustrate the methodology with a concrete example we focused on radiative transfer (RT) effects that modify the angular dependence of tracer clustering.
We illustrated that such effects may be especially problematic for inference of the growth of structure from the clustering of Lyman-$\alpha$ emitters. 
Our strategy is then applicable if these galaxies are observed concurrently with another population of galaxies that are expected to be free of this effect, such as Lyman-break galaxies.
This application is particularly timely, as early surveys of LAE clustering from DESI are now available \cite{Ebina2026:lae_desi} 
and will set the stage for the next decade of large-scale structure cosmology with these tracers from Stage-V spectroscopic surveys.
However, our simple toolbox is equally applicable to other systematic errors that may arise in 3D LSS clustering data beyond just mismodeled RT effects. 

While we worked here in the simple context of Gaussian random fields appropriate for large-scale tracer power spectra, this method may be generalized to the case of a more ambitious model that captures quasi-linear scales.
For example, applying the effective field theory + large-scale bias expansion model for tracer clustering \cite{Sullivan:2025_mtng_astrid_highz,Ebina2024J:forecast_laelbg} to these high-redshift galaxies could result in more optimal use of sample-variance cancellation.
The demonstrated success of this model in capturing these galaxies at the field-level suggests that such expanded modeling would be particularly fruitful (see Appendix~\ref{sec:perr}).
Furthermore, the methods here could be extended in this context to simultaneously infer the matter density at the field level along with the large-scale bias and stochastic parameters, a natural application for perturbative field-level modeling of astrophysical tracers such as galaxies and the Lyman-$\alpha$ forest \cite{int_bisp_lyaf,deBelsunce2026:fl_prl,Ivanov2024:LyaF_eft,Desjacques2018,Chudaykin2025:lyaf_pt_xcorr,Chen2021:Lyaf_pt,McDonald2003:Lyaf,Seljak2012:lyaf}.
For the Lyman-$\alpha$ forest, a similar method could be used to infer the value of the anisotropic linear bias $b_\eta$, exactly as described here.
We leave these exciting directions for future work.

\begin{acknowledgments}
JMS acknowledges that support for this work was provided by The Brinson Foundation through a Brinson Prize.
MW is supported by the DOE Office of Science under DE-SC0025523.
This work was performed in part at Aspen Center for Physics, which is supported by National Science Foundation grant PHY-2210452.
We thank Mikhail M. Ivanov for productive discussions during the initial stages of this project and for assistance generating Fig.~\ref{fig:perr_plot}.
We thank Arjun Dey for a useful discussion.
We thank the anonymous referee for useful comments.
This material is based upon work supported by the U.S. Department of Energy (DOE), Office of Science, Office of High-Energy Physics, under Contract No. DE–AC02–05CH11231, and by the National Energy Research Scientific Computing Center, a DOE Office of Science User Facility under the same contract.
\end{acknowledgments}

\appendix

\section{LAE \& LBG cross-stochasticity \label{sec:perr}}

\begin{figure}
    \centering
    \includegraphics[width=0.5\textwidth]{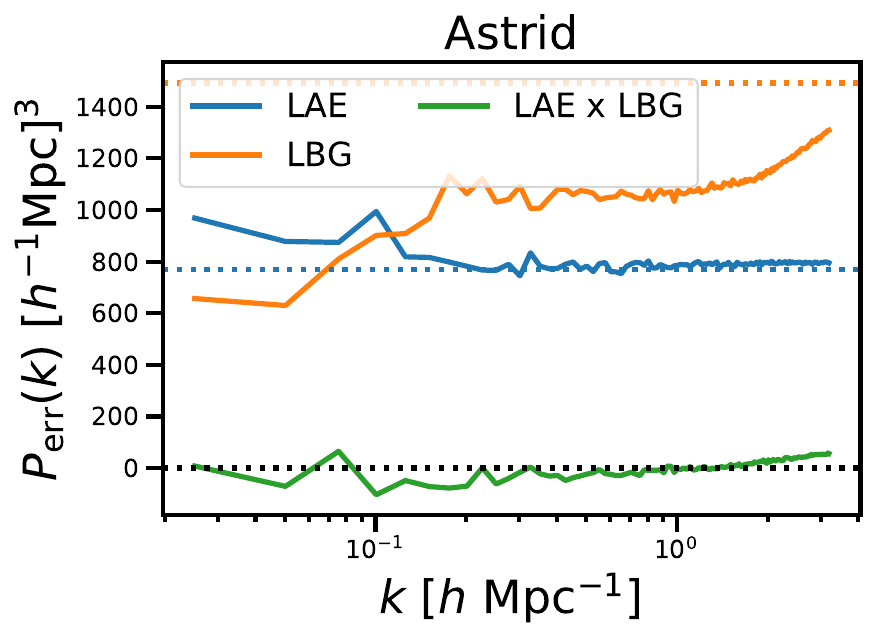}
    \caption{\textit{Stochasticity of LAEs and LBGs:} 
    The cross-stochasticity (green) as estimated from the real-space cross-error power spectrum $P_{\epsilon_1,\epsilon_2}(k)$ of the residual fields $\epsilon_1 \equiv \delta_{\mathrm{sim,LBG}}-\delta_{\mathrm{EFT,LBG}}$, and $\epsilon_2 \equiv \delta_{\mathrm{sim,LAE}}-\delta_{\mathrm{EFT,LAE}}$ computed from Ref.~\cite{Sullivan:2025_mtng_astrid_highz}.
    The cross-stochasticity (green) of the two simulated tracer populations is very nearly zero - its amplitude is suppressed by more than an order of magnitude relative to the auto-stochasticity (LAEs [blue], LBGs [orange]).
    The LBGs exhibit sub-Poisson stochasticity, as they are hosted in more massive halos than the LAEs \cite{Baldauf:2013_halo_stoch}.
    Notably, the cross-stochasticity is suppressed by an order of magnitude relative to the LAE auto-stochasticity.
    }
    \label{fig:perr_plot}
\end{figure}

The cross-stochasticity, which we denote $\bar{n}_{12}^{-1}$, is a critical component of multi-tracer forecasts.
If the cross-stochasticity were, through some mechanism, to be of a similar size to the auto-stochasticity this would significantly affect constraints, leading to lower
uncertainties. 
In this brief Appendix, we show that the cross-stochasticity is expected to be very small in the setting considered in this paper.

Figure~\ref{fig:perr_plot} shows the error power spectrum $P_{\rm{err}}$ in real space for simulated LBGs and LAEs in the Astrid simulation at $z=3$.
Specifically, these are error power spectra computed from the EFT model fits of Ref.~\cite{Sullivan:2025_mtng_astrid_highz} for the ODIN and CARS LAE and LBG samples, respectively.
Notably, the cross stochasticity (green curve) is suppressed by more than an order of magnitude with respect to the auto-stochasticities of the two tracers (blue, orange).
This justifies approximating the cross-stochasticity as very small, as we do in Section~\ref{sec:orientation}, for multi-tracer setups involving LAE-LBG cross correlation.

The smallness of the cross noise can be understood as due to the lack of overlap between the two samples under the selections applied in ref.~\cite{Sullivan:2025_mtng_astrid_highz}, see their Fig.~1.
In particular LAEs generally trace much smaller halos than LBGs (under reasonable observational selections for LBGs), and the 
expectation for cross-stochasticity of halos in widely separated mass bins is that it will be small \cite{Baldauf:2013_halo_stoch}. 
The fact that cross-stochasticity can be well approximated to be zero for the scenario we consider in this paper means that the simplified expressions presented in this limit in the main text can be taken quite seriously.

\bibliographystyle{JHEP}
\bibliography{bib}

\end{document}